\documentclass[fleqn,10pt]{wlscirep}
\usepackage[utf8]{inputenc}
\usepackage[T1]{fontenc}
\usepackage{amsmath,amssymb,amsthm}
\usepackage{graphicx}
\usepackage{hyperref}
\usepackage{soul}
\hypersetup{colorlinks=true, citecolor=orange, urlcolor=blue, linkcolor=magenta}
\usepackage[thinc]{esdiff}
\usepackage{comment}

\newcommand{\E}{\mathrm{E}}

\newcommand{\mean}[1]{\left\langle#1\right\rangle}

\title{The Fitness-Corrected Block Model, or how to create maximum-entropy data-driven spatial social networks}

\author[1]{Massimo Bernaschi}
\author[1]{Alessandro Celestini}
\author[1,*]{Stefano Guarino}
\author[1]{Enrico Mastrostefano}
\author[1,2]{Fabio Saracco}

\affil[1]{Institute for Applied Computing ``Mauro Picone'', National Research Council of Italy, via dei Taurini 19, 00185 Rome, Italy}
\affil[2]{``Enrico Fermi'' Research Center (CREF), Via Panisperna 89A, 00184 Rome, Italy}

\affil[*]{s.guarino@iac.cnr.it}

\begin{abstract}
Models of networks play a major role in explaining and reproducing empirically observed patterns.
Suitable models can be used to randomize an observed network while preserving some of its features, or to generate synthetic graphs whose properties may be tuned upon the characteristics of a given population.
In the present paper, we introduce the Fitness-Corrected Block Model, an adjustable-density variation of the well-known Degree-Corrected Block Model, and we show that the proposed construction yields a maximum entropy model.
When the network is sparse, we derive an analytical expression for the degree distribution of the model that depends on just the constraints and the chosen fitness-distribution.
Our model is perfectly suited to define maximum-entropy data-driven spatial social networks, where each block identifies vertices having similar position (e.g., residence) and age, and where the expected block-to-block adjacency matrix can be inferred from the available data.
In this case, the sparse-regime approximation coincides with a phenomenological model where the probability of a link binding two individuals is directly proportional to their sociability and to the typical cohesion of their age-groups, whereas it decays as an inverse-power of their geographic distance.
We support our analytical findings through simulations of a stylized urban area.
\end{abstract}

\begin{document}

\flushbottom
\maketitle

\thispagestyle{empty}

\section*{Introduction}

The definition of a suitable data-driven spatial social network model is a widely studied problem in computational social sciences.
Various dynamic processes (e.g., diseases' spread) can be represented on such networks, and the topology of the network has a direct impact on the evolution of the process.
Having general models for social interactions, based on available data and capable of reconstructing stylized facts known from the literature, is of utmost importance to prevent from reaching conclusions biased by incorrect or ill-defined assumptions.
With widely available survey and census data, it is now possible to generate synthetic geo-localized populations, stratified by age and organized into households. However, there is no equally direct way to accurately model interpersonal relationships.

Many real social networks exhibit some form of group mixing, driven by the tendency of individuals to socialize with their peers~\cite{mcpherson2001birds}. 
This property can be reproduced using the so-called Stochastic Block Model (SBM), risen to prominence as a way to generate networks with a known community structure~\cite{karrer2011stochastic,peixoto2014hierarchical}.
In the SBM, the vertex set is partitioned into disjoint \textit{blocks} and the probability of an edge between two nodes depends on the blocks to which the two nodes belong.
In the original formulation of the SBM, all vertices belonging to the same block are indistinguishable, so that the degree distribution within each block tends to be Poisson-like for large graphs~\cite{karrer2011stochastic}.
To produce a more realistic network, a few extensions to the model have been proposed in the literature, including the Degree-Corrected Block-Model (DCBM)~\cite{karrer2011stochastic} and its maximum-entropy version~\cite{Fronczak2013}.
These models are, in some sense, the SBM-equivalent of the well-known configuration model. They are based on enforcing both the desired group mixing and a target degree-sequence, either exactly or in expectation.  
If $L_{IJ}$ is the number of links between blocks $I$ and $J$ -- or twice that number, if $I=J$ -- and $\deg_i$ is the degree of node $v_i$, the maximum-entropy DCBM works by imposing that $\mean{L_{IJ}}=K_{IJ}$ and $\mean{\deg_i}=k_i$ for suitable constants $K_{IJ}$ and $k_i$.
The internal consistency of the model requires $\sum_J K_{IJ}=\sum_{i\in I} k_i$ for all $I$, and the density of the network is fixed equal to $\frac{\sum_{I,J} K_{IJ}}{N(N-1)}$, where $N$ is the size of the network.

In this paper, we define and analyze the Fitness-Corrected Block Model (FCBM), a variation of the DCBM where the network density $p$ is a configuration parameter.
In the FCBM the block-level mixing is specified in terms of a matrix of edge-densities $\Delta$ -- as in the original SBM -- whereas a sequence $\vec{f}$ of vertex intrinsic fitness values~\cite{caldarelli2002scale, servedio2004vertex} measures the propensity of each vertex to establish links and can be used to enforce the desired intra-block heterogeneity.
In the DCBM the constants $K_{IJ}$ and $k_i$, that bound, respectively, $\mean{L_{IJ}}$ and $\mean{\deg_i}$, need to be known explicitly.
The DCBM was in fact conceived as an instrument to randomize an observed graph while preserving some of its features.
In the FCBM, instead, $K_{IJ}$ and $k_i$ are determined based on $p$, $\Delta$ and $\vec{f}$, making the FCBM a suitable model for generating random graphs whose properties may be tuned upon the characteristics of a given population or set of entities.

To the purpose of having a model that is maximally random, the FCBM is defined following the approach first presented in Ref.~\cite{Fronczak2013} for the DCBM, and later clarified and generalized in Ref.~\cite{Cimini2019}. 
In a nutshell, the approach consists in: (i) the definition of an ensemble of networks, each with the same number of nodes, but with all possible configurations of links; (ii) the constrained maximization of the entropy associated to the network ensemble, via the method of Lagrangian multipliers.
By imposing the conditions $\mean{L_{IJ}}=K_{IJ}$ and $\mean{\deg_i}=k_i$, for all $I$, $J$, $i$, the probability per graph in the ensemble factorises in terms of probabilities per link.
The numerical value of the Lagrangian multipliers has to be calculated solving the system of nonlinear equations given by the model constraints.
The resolution of this system might become expensive when the network is large and, to the best of our knowledge, it is not implemented in any publicly available software library.
For the present paper, we implemented a parallel version of the solver for the maximum-entropy DCBM, written in C making use of the Intel MKL scientific library and the OpenMP API, which can also be used to solve our FCBM.
The solver is released as open-source software at 
\href{https://gitlab.com/cranic-group/dcbm\_solver}{https://gitlab.com/cranic-group/dcbm\_solver}.

A known sparse-regime approximation for the edge probability of the DCBM implies that, in the sparse FCBM, the probability of a link binding two individuals $i$ and $j$ is directly proportional to their sociability and to the cohesion of their blocks, i.e., $p_{ij}\propto p f_i f_j \Delta_{I_iJ_j}$ for all $i\in I$ and $j\in J$.
Under the sparse-regime approximation, the maximum entropy condition leads to a system of equations that admits a closed-form solution.
We make use of this approximate solution to find two closed-form estimates for the degree distribution $p_k$ of the FCBM -- one more accurate, the other neater.
These two estimates put in direct relation $p_k$ with the fitness distribution $p_f$, showing that the degree distribution of the FCBM, albeit not known a priori, can be essentially controlled through the model's parameters.  
In particular, if $p_f$ follows a power-law, lognormal or exponential distribution, then the same holds, approximately, for $p_k$.

Among the many possible applications, the FCBM is especially well suited for generating a data-driven social network of geo-referenced and age-stratified individuals.
The partition of the population into blocks can be obtained by grouping the individuals having similar position (e.g., residence) and age, whereas the expected block-to-block edge-density matrix $\Delta$ can be calibrated based on survey data that quantify the dependence of contact frequencies upon geographic and socio-demographic factors. 
Finally, the fitness vector $\vec{f}$ may be drawn from a suitable probability distribution, modelled upon measurable features such as wealth, employment, or mobility.
In this context, the sparse FCBM is well approximated by the phenomenological model presented in Refs.~\cite{guarino2021model,guarino2021inferring}.

To show how the FCBM can be used in practice, and to provide empirical support to our analytical findings, we generate a set of synthetic social networks for a stylized urban population distributed on a disk of radius 2.5Km.
We use the SOCRATES~\cite{Willem2020,verelst2021socrates} tool to extract age-based social mixing patterns, and we embrace the widely-accepted assumption that an inverse-power-law relation binds the distance between two individuals and the frequency of their social interactions~\cite{liben2005geographic, illenberger2013role}. 
We generate instances of our FCBM using either the exact model or its sparse-regime approximation, varying both the spatial density of the individuals in the territory and the fitness distribution.
We show that the empirical degree distribution obtained with all considered configurations is in sharp agreement with the analytical estimates.
The implementation of the FCBM is publicly available as part of the Urban Social Networks (USN) framework at \href{https://gitlab.com/cranic-group/usn}{https://gitlab.com/cranic-group/usn}.

\section*{Contributions and Results}

In the following, we summarize the main contributions and present the main analytical and experimental results of this paper.
For all methodological details, we refer the reader to the Methods section.

\subsection*{The Fitness Corrected Block Model}

We propose the Fitness-Corrected Block Model (FCBM), a new maximum-entropy model for modular networks, parameterized by the network density $p\in(0,1)$, the block-wise mixing structure $\Delta$ -- a symmetric matrix such that $\sum_{I,J}\Delta_{I,J}=2$ -- and the vertex-intrinsic fitness $\vec{f}$ -- a vector that controls the tendency of each vertex to establish links with other vertices.

Formally, let $V$ be a vertex set of size $N$ partitioned into $n$ blocks $\{B_I\}_{I=0}^{n-1}$.
The FCBM is defined as the maximum entropy probability distribution $P$, over all networks having vertex set $V$, fulfilling the following two conditions:
\begin{align}
    \label{eq:FCBM_blocks_main}
    \mean{L_{IJ}}_P &= p\binom{N}{2}\Delta_{IJ} \quad \text{for all } I,J\\
    \label{eq:FCBM_degree_main}
    \mean{\deg_i}_P &=  p\binom{N}{2} \frac{f_i}{\sum_{u\in I_i} f_u} \sum_J \Delta_{I_iJ} \quad \text{for all } i
\end{align}
where: $L_{IJ}$ is the number of edges between $B_I$ and $B_J$; $\deg_i$, $f_i$ and $I_i$ are the degree, fitness and pertaining block of vertex $v_i$; $\mean{\cdot}_P$ denotes the expected value with respect to $P$.

The FCBM can be seen as a generalization of the Degree-Corrected Block Model (DCBM).
However, contrarily to the DCBM, we show that the FCBM is consistent for any choice of the configuration parameters, making it suitable for generating random graphs with tunable topological properties.

Leveraging on the framework of entropy-based null-models, we prove that the probability $P(G)$ of generating a specific graph $G$ with the FCBM can be factorized as the product of independent edge probabilities, namely $P(G) = \prod_{i,j} p_{ij}$, where $p_{ij}$ is the probability of an edge between $v_i$ and $v_j$.
The edge probabilities can be recovered solving a system of non-linear equations obtained from (\ref{eq:FCBM_blocks_main}) and (\ref{eq:FCBM_degree_main}).

\subsection*{Efficient solver for FCBM/DCBM system of equations}

The system of nonlinear equations, needed to explicitly calculate edge probabilities, becomes computationally expensive as the size of the network increases.
Already with a few thousand nodes, a straightforward implementation may be too slow to be used in practical applications.
We implemented a parallel C-program that efficiently solves this system, as well as the analogous system arising from the DCBM.
The solver follows the Sequential Quadratic Programming approach presented in Ref.~\cite{Vallarano2021}, using Newton's method for the Hessian approximation.
To the best of our knowledge, this is the first publicly available solver for the DCBM.
The source code is publicly released as open-source software at \href{https://gitlab.com/cranic-group/dcbm\_solver}{https://gitlab.com/cranic-group/dcbm\_solver}.

\subsection*{Properties of the FCBM}

The fitness sequence $\vec{f}$ guarantees intra-block heterogeneity.
In fact, (\ref{eq:FCBM_degree_main}) can be rewritten as 
\[
\mean{\deg_i}_P =  \frac{f_i}{\sum_{u\in I_i} f_u} \mean{\deg_I}_P
\]
where $\mean{\deg_I}_P$ is the expected total degree of block $B_I$, i.e., the total number of edges incident to $B_I$.

When the network is sparse, we show that the system from which all $p_{ij}$'s must be derived admits a closed-form approximate solution.
The sparse-regime approximation allows to estimate the expected degree distribution of the sampled graph based on the fitness distribution $p_f$.
We derive two estimates for the degree distribution $p_k$ of the network.
The first estimate reads
\begin{equation}\label{eq:deg_dist_1_main}
p_k(k) \approx \frac{\mean{f}_{p_f}}{N} \sum_I
p_f\left(k \frac{\mean{f}_{p_f}}{\mu_I}\right) \frac{N_I}{\mu_I}
\end{equation}
where $\mu_I$ is the expected average degree of the vertices in $B_I$.
The second estimate, less accurate but easier to interpret and use in practice, reads
\begin{equation}\label{eq:deg_dist_2_main}
p_k(k) \approx
p_f\left(k \frac{\mean{f}_{p_f}}{\mu}\right) \frac{\mean{f}_{p_f}}{\mu}
\end{equation}
where $\mu$ is the average degree of the network.
The obtained analytical expressions for $p_k$ have a very desirable property: for many choices of $p_f$ -- including power-law, lognormal or exponential distributions -- $p_k$ essentially has the same ``shape'' of $p_f$.

\subsection*{Data-driven FCBM for spatial social networks}

We propose an application for our FCBM as a maximum-entropy model for data-driven spatial social networks. 
In particular, we envision its application to generate geographic networks informed by census data, contact surveys, and geospatial data. Indeed, the phenomenological model described at the end of this section has already been employed to develop a realistic social network at the urban scale \cite{guarino2021inferring} and to study the spread of an epidemic process \cite{guarino2021data,celestini2021epidemics,celestini2022epidemic}, on it. 

Let the vertex set $V$ describe an age-stratified population of $N$ individuals living in a territory tessellated into square tiles of side $l$.
Each $v_i$ is thus characterized by two data-driven discrete attributes: its tile of residence $t_i\in T$, that is, the discretized position of $v_i$ in the territory, and its age-group $g_i\in \Gamma$. 
These two attributes induce a partition of the population into $n=|T|\cdot |\Gamma|$ blocks $\{B_I\}_{i=0}^{n-1}$, with $v_i\in B_I=(t_I,g_I)$ if and only if $t_i=t_I$ and $g_i=g_I$.

To define a data-driven block-wise mixing matrix $\Delta$, we observe that:
\begin{itemize}
    \item Social mixing patterns, derived from heterogeneous data sources such as surveys, cell phones or wearable sensors~\cite{Mossong-2008,eagle2009inferring,klepac2020contacts}, can be used to reconstruct a data-driven age-based mixing matrix $S$~\cite{guarino2021inferring}, whose $s_{IJ}$ element measures the tendency of age groups $g_I$ and $g_J$ to socialize with each other.
    \item The frequency of social relations between individuals living in $t_I$ and $t_J$ is generally assumed to decay as $d_{IJ}^{-\beta}$, where $d_{IJ}$ is the normalized (geographic or euclidean) distance between tiles $t_I$ and $t_J$, and the exponent $\beta>0$ depends on the type of relation and on the extension of the territory~\cite{liben2005geographic, illenberger2013role}.
\end{itemize}
This leads to:
\begin{equation}\label{eq:Delta}
    \Delta_{IJ} = \frac{d_{IJ}^{-\beta}s_{IJ}}{\sum_{O\leq Q}d_{OQ}^{-\beta}s_{OQ}}
\end{equation}

Finally, we extract the fitness vector $\vec{f}$ from a suitable distribution $p_f$, set the density parameter $p\in(0,1)$ and, for all pairs $i,j$, compute the edge probability $p_{ij}$ as prescribed by the FCBM.
As a result, the graph sampling probability $P$ guarantees that: (i) the expected number of links between $B_I$ and $B_J$ is proportional to $s_{I,J}$ and decays as $d_{I,J}^{-\beta}$; (ii) the expected degree of $v_i$ is proportional to $f_i$ and to the expected total degree of block $I_i$.

In this case, the sparse-regime approximation yields
\begin{equation}\label{eq:FCBM_data_main}
p_{ij} \approx p \binom{N}{2} \frac{f_i}{\sum_{u\in I_i} f_u} \frac{f_j}{\sum_{w\in J_j} f_w}\frac{d_{I_iJ_j}^{-\beta}s_{I_iJ_j}}{\sum_{I\leq J}d_{IJ}^{-\beta}s_{IJ}}
\end{equation}
Expression (\ref{eq:FCBM_data_main}) defines a phenomenological model, analogous to the one presented in~\cite{guarino2021inferring}, where the probability of two individuals being connected is proportional to their sociability and to the cohesion of their age-groups, while decaying as a power of their distance.
Clearly, the estimates obtained in (\ref{eq:deg_dist_1_main}) and (\ref{eq:deg_dist_2_main}) for the degree distribution stay valid in the data-driven model.
If the network is sufficiently sparse and the population of all tiles/groups is sufficiently large, the degree distribution of the sampled graph $G$ is controlled by the available data and by the chosen fitness distribution $p_f$.

\subsection*{Experimental Analysis}

We implemented both the exact FCBM and its sparse-regime approximation.
The code is released, as open-source software, as part of the Urban Social Networks (USN) framework at \href{https://gitlab.com/cranic-group/usn}{https://gitlab.com/cranic-group/usn}.

We used our data-driven FCBM to generate instances of a spatial social network for a stylized city of 10K inhabitants living in disk of radius 2.5Km.
We set $p$ so that the average degree of the network is $\mu=25$, we set $\beta=1$, we used age-density data for Italy as released by the Italian National Institute of Statistics (ISTAT), and we extracted the matrix $S$ from data released by the POLYMOD project~\cite{Mossong2008}.
We considered three possible spatial densities -- uniform, and increasing or decreasing with the distance from the disk's center -- and three possible fitness distributions -- pareto, lognormal and exponential.
For all nine combinations, we generated 10 independent graph instances. 

\begin{figure}[htbp]
    \centering
    \includegraphics[width=.9\textwidth]{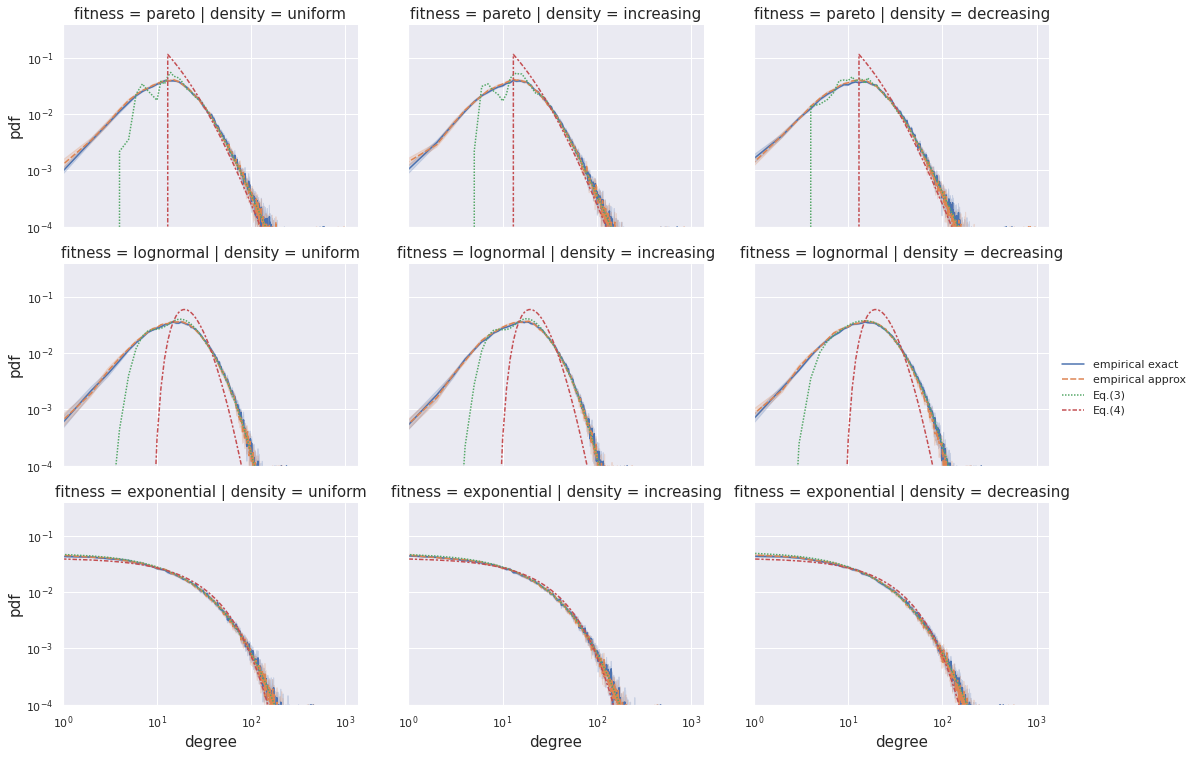}
    \caption{The degree distribution for an ideal city (circular, 10K inhabitants) for 3 different densities (uniform, radial increasing, radial decreasing) and 3 different fitness distributions (powerlaw, lognormal, exponential).}
    \label{fig:deg_dis}
\end{figure}

In Figure~\ref{fig:deg_dis} we show the empirical degree distribution, averaged over the 10 instances of the FCBM for each configuration, considering both the exact model and the sparse-regime approximation.
We also show the two estimates (\ref{eq:deg_dist_1_main}) and (\ref{eq:deg_dist_2_main}).
The plots confirm that the approximation is sound and that the two estimates can be safely used in practice -- at least, for sparse networks -- with (\ref{eq:deg_dist_1_main}) working especially well in all cases -- except, possibly, for very small degrees.

\newpage

\section*{Discussion}

Developing realistic network models for social interactions is of paramount importance to understand the underlying mechanisms that lead to the observed features of such networks and to study all dynamic processes, such as disease spreading, that are strongly influenced by the network topology.
Extreme care must be taken to avoid that any bias is unintentionally injected in the model.
For this reason, maximum-entropy models are extensively used in network analysis, either as null models, or to generate synthetic networks that have specific characteristics, but are otherwise maximally random.

In this paper, we introduced the maximum-entropy Fitness-Corrected Block Model (FCBM), an adjustable-density model for modular and heterogeneous networks, whose block-wise mixing pattern is known in expectation.
The model has a general and flexible formulation, but it was designed with a key application in mind: providing a working tool for building synthetic social networks informed with census, survey and geospatial data.
Publicly available spatial density and demographic data are in fact necessarily discrete, thus inducing a partition of the population into blocks of agents who belong to the same age-group and live in the same area.
The expected block-to-block edge-density can be estimated based on empirical findings that quantify the dependence of contact frequencies upon geographic and demographic features.
However, both the density and the degree of heterogeneity of real-world social networks depend on the considered type of interpersonal relations and are rarely explicitly known beforehand.
Contrarily to other block-models in the literature, the FCBM makes both these network features adjustable.
In particular, the desired intra-block heterogeneity can be enforced through a vertex-intrinsic social fitness, possibly modelled upon observable population-level variables, such as wealth, employment or mobility.

We implemented the FCBM and made it publicly available as open-source software.
The released software includes a parallel code that speeds up the computation of the most expensive part of the required maximization procedure, a step of the algorithm that is also needed in the well-known Degree-Corrected Block Model.
We tested our implementation of the FCBM by reconstructing instances of a social network connecting the individuals of a stylized city of 10K inhabitants.
The experiments allowed to verify that the exact model and its efficient sparse-regime approximation yield networks with almost identical degree distributions.
We also showed and experimentally verified that, in the sparse-regime, the expected degree distribution of the output network can be estimated by two closed-form expressions.
Thanks to these two estimates, the shape of the degree distribution can be predicted based on the chosen fitness distribution.
 
In the next future, we plan to use the FCBM -- and, possibly, a temporal extension of the model -- to simulate dynamic processes in real-world territories and understand how socio-demographic features and social habits affect the outcomes of these processes.
To this end, we will work towards gaining a better understanding of how the topological properties of the FCBM depend on the configuration parameters, with special attention paid to a set of network properties, such as the clustering coefficient and the excess degree distribution, that can be used to study percolation and diffusion dynamics on the network.



\section*{Related Work}


Under the pressure of the COVID-19 pandemic, several simulation frameworks have been developed to provide realistic descriptions of the disease spread process on different spatial and temporal scales, from a single building to complex urban areas up to a global scale \cite{kerr2021covasim,mahmood2020facs,liu2021modelling, coletti2021data,bouchnita2020hybrid, aleta2020modelling,aleta2022quantifying,chang2021mobility}. 
World-scale meta-population models and agent-based systems describing small and large areas can be informed by a variety of data sources. Census data and/or grid based population counts can be integrated to reconstruct populations that are statistically indistinguishable from real ones, including age, geographic distribution, education and wealth. The number and intensity of contacts in specific settings, such as workplaces, schools, or households, can be collected through surveys, questionnaires, diaries, and, if possible, supplemented with data obtained from digital technologies such as cell phones or wearable sensors ~\cite{Mossong-2008,eagle2009inferring,klepac2020contacts}. These data are then used to reconstruct contact matrices, individual or group schedules, and are widely used to reproduce synthetic interactions. ~\cite{del2007mixing,barrett2009generation}. 

In comparison, generative network models that can reproduce the characteristics of real-world social networks by incorporating information from data, have received much less attention. The task of inferring a realistic distribution of social ties is quite challenging, since friendship ties can only be measured for a small subset of real-world networks, and the mechanism underlying tie formation, while thoroughly studied, is still far from being fully understood.
Ideally, the models should reproduce the main features of real-world social networks, well summarized in~\cite{kertesz2021modeling}. These networks show a heavy-tailed (e.g, lognormal) degree distribution, often with a finite cutoff in agreement with Dunbar's number. The transitivity of the networks is high, compared to a random graph model, as a consequence of the well-established principle that ``friends of my friends are my friends." Moreover, they show positive assortativity by degree and \emph{type}.
By quantitatively looking at ego networks, mobile phone networks,  and online social networks we now have a better  understanding of some of their peculiar features and underline mechanisms.
Education, wellness, age and spatial proximity are regarded as critical elements in the formation of friendship bonds~\cite{palla2007quantifying,HUANG2013969}, among these, age is probably the most studied possibly due to the fact that age-related data is actually available at different spatial scales~\cite{worldpop}. 
While there is wide evidence that geographical factors alone cannot explain the structure of real-world spatial social networks~\cite{scellato2011socio,liben2005geographic,herrera2015anatomy}, the dependence of friendship on distance is widely assumed to follow an inverse power-law with exponent $\beta\in[0.5, 2]$~\cite{lambiotte2008geographical,scellato2011socio,herrera2015anatomy,onnela2011geographic,liben2005geographic,illenberger2013role,walsh2011spatial,buchel2020cities} -- and this surprisingly holds even for online relationships~\cite{goldenberg2009distance}. In particular, $\beta<1$ seems to work better for short range contacts ($<20km$)~\cite{illenberger2013role} and for urban networks~\cite{walsh2011spatial},
in line with sociological studies~\cite{krackhardt2003strength}, but, contrary to other real-world networks~\cite{bernaschi2019spiders,newman2010networks}, such networks do not present very large hubs~\cite{onnela2007analysis,lambiotte2008geographical,herrera2015anatomy,liben2005geographic}.
Their degree distribution is right skewed and relatively long-tailed~\cite{herrera2015anatomy,liben2005geographic}, and it has been, at times, approximated by a power-law with a large ($5$ to $8$) exponent~\cite{onnela2007analysis,lambiotte2008geographical} or by a Lognormal distribution~\cite{illenberger2013role}.
Within cities, population density impacts on the frequency of close-range contacts, but usually not on the overall size of each person's network~\cite{buchel2020cities}.
While geographical proximity and community structure appear to be related~\cite{herrera2015anatomy,walsh2011spatial,buchel2020cities}, some authors argue that only small clusters ($<30$ members) are geographically bounded~\cite{onnela2011geographic}, whereas the large ones may span across very large areas of a city~\cite{herrera2015anatomy}.

Defining simple models that capture all of these features is not an easy task.
Models designed to mimic the scale-free degree distribution emerging in many real networks, for instance, may fail to yield the expected clustering structure~\cite{cointet2007realistic,iskhakov2020local}. 
Exponential random graphs have been shown to overcome some of these limitations \cite{robins2007recent,daraganova2012networks}.

The community structure, which evidences the presence of a kind of homophily, is a typical characteristic of real-world social networks and  Stochastic Block Models (SBMs) have been developed specifically to reproduce and study this feature. ~\cite{karrer2011stochastic,peixoto2014hierarchical}
In this type of network models, the nodes are partitioned into disjoint sets named \textit{blocks} and the probability of an edge existing between two nodes depends on the blocks to which the two nodes belong.
The SBM and its generalization have gained their success in the last decades as they can be used to discover and understand the structure of a network, as well as for clustering purposes~\cite{mccallum2007topic,zhou2006probabilistic}.

Spatial network models are often obtained by incorporating vertices into a metric space and induced constraints can determine some of the network properties ~\cite{barthelemy2011spatial}. Introducing a penalty on "long" edges, which mimics a penalty in maintaining long-distance relationships, has an impact on clusters, path lengths, degree distributions, and more ~\cite{alizadeh2017generating}.

Recently, network instances having suitable features have been generated by means of the so-called \textit{random geometric} models~\cite{krioukov2010hyperbolic,boguna2010sustaining,serrano2008self}, where the popularity and similarity of the nodes depend on their position in some \emph{latent} metric space~\cite{papadopoulos2012popularity}.
Embedding the vertices into a hyperbolic disk~\cite{krioukov2010hyperbolic} has proved a way to obtain both high clustering and heavy-tailed degree distribution.

In the present paper, we leverage on the framework of entropy-based null-models for real complex networks, revised in Ref.~\cite{Cimini2019}. Among the very first fundamental papers, the work of Park and Newman has a particular relevance~\cite{park2004statistical}: based on Jaynes' derivation of Statistical Physics from Information Theory~\cite{Jaynes1957}, they proposed a general maximum entropy approach for the randomization of complex networks. Among the extension to a different context, the main innovations of Ref.~\cite{park2004statistical} are the introduction of local constraints, as the degree sequence, and the interpretation of the general framework of Exponential Random Graphs (ERGs) in terms of maximum entropy models. The present construction was later extended to the analysis of real networks~\cite{Garlaschelli2004, squartini2011analytical}, tailoring the entropy-based model on the observed network. As a matter of fact, the various Lagrangian multipliers, introduced for the entropy maximisation in Ref.~\cite{park2004statistical} can be numerically calculated by maximising the (Log-)Likelihood associated with the real network. Such a construction represents a perfect benchmark for the analysis of real systems, since it is maximally random (due to the entropy maximisation) and tailored on the observed system (due to the Likelihood maximisation). 
It is not surprising that it has been extensively applied to the study of non trivial structural patterns of different systems, as financial and trade networks, biological systems and online social networks~\cite{Straka2018, Cimini2019}. Moreover, the general framework can be easily extended to tackle different kinds of networks, as undirected, directed, weighted, directed and weighted~\cite{Squartini2013,Mastrandrea2014,Squartini2015f}, bipartite~\cite{Saracco2015}, bipartite weighted~\cite{DiGangi2018} 
and degree corrected block models~\cite{Fronczak2013}.
Another relevant branch of research focuses on the reconstruction of networks from limited information~\cite{Squartini2018}. 
This application is of particular interest for risk assessment of financial networks, and limited information is available due to privacy concerns~\cite{Bardoscia2021}. 
Reconstruction approaches based on entropy-based null models have proven particularly effective in this context, and the maximally random nature of the framework described here is critical to avoid the introduction of bias into the predictions~\cite{Gandy2017, Ramadiah2017,Anand2018,Squartini2018,Bardoscia2021}.

\section*{Methods}

\subsection*{Formalism}
Let $\mathcal{G}$ be the ensemble of all simple graphs of $N$ vertices.
If $P$ is a probability distribution over $\mathcal{G}$, $P(G)$ is the probability of graph $G\in\mathcal{G}$, and $\mean{\cdot}_P$ denotes the expectation with respect to $P$.
The vertex set $V=\{v_i\}_{i=0}^{N-1}$ is partitioned into $n$ blocks $\{B_I\}_{I=0}^{n-1}$.
The size of block $I$ is $N_I=|B_I|$ and, for each $v_i\in V$, $I_i$ denotes the index of the block to which $v_i$ belongs.
For all pairs $I,J$, $N_{IJ}$ denotes the number of possible pairs $(i,j)$ with $v_i\in I$, $v_j\in J$ and $i\neq j$, i.e., $N_{II}=N_I(N_I-1)$ and $N_{IJ}=N_IN_J$ if $I\neq J$ -- notice that in the case of $N_{II}$ we are counting twice the number of couples, as we will do for the counts of edges in the following.

Each $G\in\mathcal{G}$ is uniquely determined by its adjacency matrix $A(G)=\{a_{ij}(G)\}_{i,j=0}^{N-1}$, where $a_{ij}(G)=1$ if edge $(i,j)\in E(G)$ and $a_{ij}(G)=0$ otherwise. 
The degree of vertex $v_i$ in $G$ is $\deg_i(G)=\sum_j a_{ij}(G)$.
The total degree of block $I$ is $\deg_I(G) = \sum_{i\in I} \deg_i(G)$ and, for all $I,J$, $L_{IJ}(G)=\sum_{i\in I}\sum_{j\in J} a_{ij}(G)$ is the number of edges between $B_I$ and $B_J$ in $G$, or, if $I=J$, twice that number.
Therefore, using the definitions, $\deg_I(G)=\sum_J L_{IJ}(G)$.
For the sake of simplicity, the dependence of these quantities on the specific graph $G$ will be often omitted in the following.

\subsection*{Maximum Entropy Degree-Corrected Block Model~\cite{Fronczak2013}}

The maximum entropy Degree-Corrected Block Model (DCBM) is defined as the maximum entropy probability distribution $P$ over $\mathcal{G}$ in which the number of links per block and the degree sequence are constrained on average, i.e.
\begin{align}
    \label{eq:DCBM_blocks}
    \mean{L_{IJ}}_P &= K_{IJ} \quad \text{for all } I,J\\
    \label{eq:DCBM_degree}
    \mean{\deg_i}_P &= k_i \quad \text{for all } i,
\end{align}
with $\sum_J K_{IJ}=\sum_{i\in I} k_i$, for all $I$.
If $H(P)$ denotes the Shannon entropy of $P$, the sought $P$ can be obtained by finding the stationary points of 
\begin{equation}
H'(P,\vec{\eta},\vec{\theta})=H(P)-C(P,\vec{\eta},\vec{\theta}) = -\mean{\ln P(G)}_P - \left(\sum_{I\leq J}\eta_{IJ}\left(\mean{L_{IJ}}_P-K_{IJ}\right) + \sum_i\theta_i\left(\mean{\deg_i}_P-k_i\right)+\alpha(\sum_G P(G)-1)\right)
\end{equation}
where $\eta_{IJ}$, $\theta_i$ and $\alpha$ are Lagrange multipliers: while, $\eta_{IJ}$ and $\theta_i$ control the conditions (\ref{eq:DCBM_blocks}) and (\ref{eq:DCBM_degree}), respectively, $\alpha$ is necessary for the normalization of the probability $P(G)$.
Since the functional derivatives with respect to $P(G)$ are $\diff{\mean{L_{IJ}}_P}{P(G)}=
\sum_{i\in I}\sum_{j\in J} a_{ij}(G)$ and $\diff{\mean{\deg_i}_P}{P(G)}=
\sum_j a_{ij}(G)$, then
\[
\diff{C(P,\vec{\eta},\vec{\theta})}{P(G)}=
\sum_{I\leq J}\eta_{IJ}\sum_{i\in I}\sum_{j\in J} a_{ij}(G) + \sum_i \theta_i \sum_j a_{ij}(G) +\alpha= \sum_{i<j}\eta_{I_iJ_j}a_{ij}(G) + \sum_{i<j} (\theta_i + \theta_j) a_{ij}(G)+\alpha
\]
This results in
\begin{equation*}
P(G)\propto \exp\left[-\sum_{i<j}(\theta_i+\theta_j+\eta_{I_iJ_j})a_{ij}(G)\right]
\end{equation*}
and the probability per graph factorises in terms of probabilities per link as
\begin{equation}\label{eq:DCBM}
    p_{ij}=\begin{cases} \dfrac{x_ix_jy_{I_iJ_j}}{1+x_ix_jy_{I_iJ_j}} &\text{if } i\neq j\\
    0 &\text{otherwise}
    \end{cases}
\end{equation}
where $x_i=e^{-\theta_i}$ and $y_{IJ}=e^{-\eta_{IJ}}$.
For all $i$ and all $I\leq J$, $x_i$ and $y_{IJ}$ can be found by solving the system of equations
\begin{align}
    \label{eq:blocks}
   \sum_{i\in I}\sum_{j\in J}p_{ij} &= K_{IJ} \quad \text{for all } I\leq J\\
    \label{eq:deg}
    \sum_j p_{ij} &= k_i \quad \text{for all } i
\end{align}

\paragraph{Sparse DCBM}


When the average degree $\mean{k}$ is small, i.e., when the network is sparse, the edge probability of the DCBM can be approximated as $p_{ij}\approx x_ix_jy_{I_iJ_j}$. This allows to rewrite (\ref{eq:blocks}) and (\ref{eq:deg}) as
\begin{align}
    \label{eq:blocks_const}
    y_{IJ} \sum_{i\in I} x_i \sum_{\substack{j\in J\\j\neq i}}x_j &= K_{IJ} \quad \text{for all } I\leq J\\
    \label{eq:deg_const}
    x_i \sum_J y_{I_iJ} \sum_{\substack{j\in J\\j\neq i}}x_j  & = k_i \quad \text{for all } i 
\end{align}

Since the constants $K$ 
and $\vec{k}$ are, by construction, bound by the relation $\sum_{J} K_{IJ} = \mean{\deg_I} = \sum_{i\in I} k_i$, (\ref{eq:blocks_const}) and (\ref{eq:deg_const}) admit the following solution
\begin{align*}
    x_i &= k_i \quad \text{for all } i\\
    y_{IJ}  &= \frac{K_{IJ}}{\left(\sum_{i\in I}k_i\right) \left(\sum_{\substack{j\in J\\j\neq i}}k_j\right)} \quad \text{for all } I\leq J
\end{align*}

\subsection*{Maximum Entropy Fitness-Corrected Block Model}

Given a scalar $p\in(0,1)$, a symmetric matrix $\Delta$ 
such that $\sum_{I,J}\Delta_{I,J}=2$, and a fitness sequence $\vec{f}$, we define the Fitness-Corrected Block Model (FCBM) as the maximum entropy model fulfilling the following two conditions:
\begin{align}
    \label{eq:FCBM_blocks}
    \mean{L_{IJ}}_P &= p\binom{N}{2}\Delta_{IJ} \quad \text{for all } I, J\\
    \label{eq:FCBM_degree}
    \mean{\deg_i}_P &= p\binom{N}{2} \frac{f_i}{\sum_{u\in I_i} f_u} \sum_J \Delta_{I_iJ} \quad \text{for all } i
\end{align}

The FCBM is a variation of the DCBM where the network density $p$ is a configuration parameter.
As in the original stochastic block model, the block-level mixing is specified in terms of a set of edge-densities $\Delta_{IJ}$, rather than a set of edge-counts $K_{IJ}$.
Similarly, the degree sequence $\vec{k}$ is replaced by a vertex intrinsic fitness sequence $\vec{f}$, in line with previous models available in the literature~\cite{caldarelli2002scale, servedio2004vertex}.
$f_i$ measures $v_i$'s propensity to establish links and $\mean{\deg_i}_P$ is set proportional to $f_i$ by a constant that depends on $I_i$, other than $p$.
By design, (\ref{eq:FCBM_blocks}) and (\ref{eq:FCBM_degree}) imply $\sum_J\mean{L_{IJ}}_P = \mean{\deg_I}_P = \sum_{i\in I} \mean{\deg_i}_P$, and (\ref{eq:FCBM_degree}) can be rewritten as \[
\mean{\deg_i}_P =  \frac{f_i}{\sum_{u\in I_i} f_u} \mean{\deg_I}_P
\]
which clarifies the role of $\vec{f}$ as an element of intra-block heterogeneity.

For fixed $\Delta$ 
and $\vec{f}$, the derivation of the maximum entropy FCBM is identical to that of the DCBM: the maximum entropy probability per graph factorizes into the probability per edge given by (\ref{eq:DCBM}) and the vectors of constants $\vec{x}$ and $\vec{y}$ can be obtained by solving the analogous of (\ref{eq:blocks}) and (\ref{eq:deg}), i.e.
\begin{align}
    \label{eq:blocks_FCBM}
    \sum_{i\in I}\sum_{j\in J}p_{ij} &= p\binom{N}{2}\Delta_{IJ} \quad \text{for all } I\leq J\\
    \label{eq:deg_FCBM}
    \sum_j p_{ij} &= p\binom{N}{2}\frac{f_i}{\sum_{u\in I_i} f_u} \sum_J \Delta_{I_iJ} \quad \text{for all } i
\end{align}

\paragraph{Sparse FCBM}
In the sparse-regime, i.e, when $p\ll 1$, using $p_{ij}\approx x_ix_jy_{I_iJ_j}$ yields the following approximate solution to (\ref{eq:blocks_FCBM}) and (\ref{eq:deg_FCBM}): 
\begin{align*}
    x_i &= p\binom{N}{2}\frac{f_i}{\sum_{u\in I_i} f_u} \sum_J \Delta_{I_iJ} = \frac{f_i}{\sum_{u\in I_i} f_u} \mean{\deg_I}_P \quad \text{for all } i\\
    y_{IJ}  &= \frac{p\binom{N}{2}\Delta_{IJ}}{ 
    p\binom{N}{2}\left(\sum_O \Delta_{IO} \right)
    p\binom{N}{2}\left(\sum_O \Delta_{OJ} \right)} = \frac{p\binom{N}{2}\Delta_{IJ}}{\mean{\deg_I}_P \mean{\deg_J}_P} \quad \text{for all } I\leq J
\end{align*}
In this regime, the probability per edge can thus be rewritten as
\begin{equation}\label{eq:FCBM_main}
p_{ij} \approx p\binom{N}{2} \frac{f_i}{\sum_{u\in I_i} f_u} \frac{f_j}{\sum_{w\in J_j} f_w} \Delta_{I_iJ_j}
\end{equation}

\paragraph{Degree distribution for the sparse FCBM}
Let us assume that each fitness value $f_i$ is drawn from a suitable distribution $p_f$.
The sparse-regime approximation allows to estimate the expected degree distribution of the sampled graph $G$ with respect to both $p_f$ and the graph sampling probability $P$.
If $N_{I_i}$ is large enough, we have $\sum_{u\in I_i }f_u \approx \mean{f}_{p_f} N_{I_i}$.
Now, following the approach used in ~\cite{caldarelli2002scale}, we have
\begin{align}\label{eq:exp_deg_cond}
k(f,I) = \mean{\deg_i\bigm\vert f_i=f,I_i=I}_{p_f,P} 
\approx 
\dfrac{f}{\mean{f}_{p_f}N_{I}}\mean{\deg_I}_P = \dfrac{f}{\mean{f}_{p_f}} \mu_I 
\end{align}
where $\mu_I = \frac{\mean{\deg_I}_P}{N_I} = \mean{\deg_i \bigm\vert I_i=I}_P$ is the expected average degree of the vertices in $I$.
Equation (\ref{eq:exp_deg_cond}) can be inverted leading to estimate
\[
f(k,I) \approx k \frac{\mean{f}_{p_f}}{\mu_I}
\]
so that
\begin{equation}\label{eq:deg_dist_I}
p_k(k,I) = \Pr[k(f)=k\bigm\vert I] = \Pr[f(k)=f\bigm\vert I]\diff{f(k)}{k} \approx
p_f\left(k \frac{\mean{f}_{p_f}}{\mu_I}\right) \frac{\mean{f}_{p_f}}{\mu_I}
\end{equation}
and, hence,
\begin{equation}\label{eq:deg_dist_1}
p_k(k) = \Pr[k(f)=k] 
= \sum_I p_k(k,I)\frac{N_I}{N} 
\approx \sum_I
p_f\left(k \frac{\mean{f}_{p_f}}{\mu_I}\right) \frac{\mean{f}_{p_f}}{\mu_I} \frac{N_I}{N}
= \frac{\mean{f}_{p_f}}{N} \sum_I
p_f\left(k \frac{\mean{f}_{p_f}}{\mu_I}\right) \frac{N_I}{\mu_I}
\end{equation}
A slightly less accurate, yet much simpler, approximation can be obtained computing first
\begin{equation}\label{eq:exp_deg}
    k(f) 
    = \mean{\deg_i\bigm\vert f_i=f}_{p_f,P}
    = \sum_I k(f,I) \frac{N_I}{N} 
    \approx \frac{f}{\mean{f}_{p_f}} \sum_I \mu_I \frac{N_I}{N} =
    \frac{f}{\mean{f}_{p_f}} \mu
\end{equation}
where $\mu=\sum_I \mu_I \frac{N_I}{N}$ is the average degree of the network.
Then, (\ref{eq:exp_deg}) can be inverted as
\[
f(k) \approx k \frac{\mean{f}_{p_f}}{\mu}
\]
yielding
\begin{equation}\label{eq:deg_dist_2}
p_k(k) = \Pr[k(f)=k] = \Pr[f(k)=f]\diff{f(k)}{k} \approx
p_f\left(k \frac{\mean{f}_{p_f}}{\mu}\right) \frac{\mean{f}_{p_f}}{\mu}
\end{equation}
In many cases, $p_k$ belongs to the same family of probability distributions of $p_f$: e.g., if $p_f$ follows a power-law, lognormal or exponential distribution, then the same holds for $p_k$.

\subsection*{Data-driven FCBM for spatial social networks}

Our FCBM can be easily tuned upon real data and empirical findings to produce instances of a maximum entropy spatial social network.
On one hand, the local density and demographic profile of the population are generally available in the form of discrete, geographically located (e.g., residents in 500m$\times$500m tiles) and/or age-stratified (e.g., 0-5 years old) population segments.
These data naturally induce a partition of the population into blocks.
On the other hand, intra-block population heterogeneity can be controlled by a vertex-related social ﬁtness, possibly modelled upon measurable features such as wealth, employment or mobility.

Formally, let the vertex set $V$ describe an age-stratified population of $N$ individuals living in a territory tessellated into square tiles of side $l$.
Each $v_i$ is characterized by two data-driven discrete attributes: its tile of residence $t_i\in T$, that is, the discretized position of $v_i$ in the territory, and its age-group $g_i\in \Gamma$. 
$t_i$ and $g_i$ may either be directly available -- in the case of a real population -- or be drawn, respectively, from given spatial density $p_t$ and age-distribution $p_g$ -- in the case of a synthetic population.
These two attributes induce a partition of the population into $n=|T|\cdot |\Gamma|$ blocks $\{B_I\}_{i=0}^{n-1}$, with $v_i\in B_I=(t_I,g_I)$ if and only if $t_i=t_I$ and $g_i=g_I$.
We embrace the widely-acknowledged assumption that an inverse-power-law relation binds the distance $d_{IJ}$ and the frequency of social relations between individuals living in $t_I$ and $t_J$~\cite{liben2005geographic, illenberger2013role}.
For all pairs of blocks $I,J$, we thus define the edge-density 
\[
\Delta_{IJ} = \frac{d_{IJ}^{-\beta}s_{IJ}}{\sum_{O\leq Q}d_{OQ}^{-\beta}s_{OQ}}
\]
where
\begin{itemize}
    \item $d_{IJ}$ is the normalized (geographic or euclidean) distance between tiles $t_I$ and $t_J$; the normalization is obtained through a division by $\frac{l}{2}$. We set $d_{II}=1$, so that the distance between individuals in the same tile is half the distance of individuals living in neighboring tiles.
    \item $\beta>0$ is a configuration parameter.
    \item $s_{IJ}$ measures the tendency of age groups $g_I$ and $g_J$ to socialize with each other; such a $|\Gamma|\times |\Gamma|$ symmetric age-based social mixing matrix $S$ can be obtained, by imposing reciprocity and normalizing, from a suitable data-driven contact matrix~\cite{guarino2021inferring}.
\end{itemize}
Finally, we extract the fitness vector $\vec{f}$, set the density parameter $p\in(0,1)$ and, for all pairs $i,j$, compute the edge probability $p_{ij}$ as described for the FCBM.
As a result, the graph sampling probability $P$ guarantees that: (i) the expected number of links between $B_I$ and $B_J$ is proportional to $s_{I,J}$ and decays as $d_{I,J}^{-\beta}$; (ii) the expected degree of $v_i$ is proportional to $f_i$ and to the expected total degree of block $I_i$.

In this case, the sparse-regime approximation yields
\begin{equation}\label{eq:FCBM_data}
p_{ij} \approx p \binom{N}{2} \frac{f_i}{\sum_{u\in I_i} f_u} \frac{f_j}{\sum_{w\in J_j} f_w}\frac{d_{I_iJ_j}^{-\beta}s_{I_iJ_j}}{\sum_{I\leq J}d_{IJ}^{-\beta}s_{IJ}}
\end{equation}
Expression (\ref{eq:FCBM_data}) defines a phenomenological model, where the probability of two individuals being connected is proportional to their sociability and to the cohesion of their age-groups, while decaying as a power of their distance.
Clearly, the estimates obtained in (\ref{eq:deg_dist_1}) and (\ref{eq:deg_dist_2}) for the degree distribution stay valid in the data-driven model.
If the network is sufficiently sparse and the population of all tiles/groups is sufficiently large, the degree distribution of the sampled graph $G$ is controlled by the available data and by the chosen fitness distribution $p_f$.

\bibliography{synth_pop.bib}


\section*{Acknowledgements}
This work was supported in part by the Project ``CARES: Context-Aware Realistic Epidemic Simulator'', funded by the Italian Ministry of Research under the FISR 2020 programme.
The Ministry of Research had no role in the
design of the study and collection, analysis, and interpretation of data and in writing the manuscript.
Any opinion, finding, and conclusions expressed in this paper only reflect the views of the authors.


\section*{Author contributions statement}
All authors designed the study.
S.G. and F.S. developed the methods.
A.C. and E.M. acquired the data.
M.B., A.C, S.G and E.M. created the software.
S.G. interpreted the experimental results and wrote most of the paper.
All authors contributed to the writing and the revision of the article, and they all read and approved the final manuscript.


\section*{Competing interests}
The authors declare no competing interests.

\section*{Data availability statement}
All code and data used in this paper are available at the following public repositories: \href{https://gitlab.com/cranic-group/usn}{https://gitlab.com/cranic-group/usn} and 
\href{https://gitlab.com/cranic-group/dcbm\_solver}{https://gitlab.com/cranic-group/dcbm\_solver}.

\section*{Additional information}
\textbf{Correspondence} and requests for materials should be addressed to S.G.

\end{document}